\documentclass[11pt]{article}
\usepackage{amsfonts}
\usepackage{amssymb}
\usepackage{amsmath,amssymb}
\usepackage{bbm}

\def\slasha#1{\setbox0=\hbox{$#1$}#1\hskip-\wd0\hbox to\wd0{\hss\sl/\/\hss}}

\def\periodb#1{\setbox0=\hbox{$#1$}#1\hskip-\wd0\hbox to\wd0{-}}

\def\sfrac#1#2{{\textstyle\frac{#1}{#2}}}
\newcommand{\unit}{\mathbbm{1}}   % identity map/matrix

\newcommand{\frg}{\mathfrak{g}}
\newcommand{\frh}{\mathfrak{h}}
\newcommand{\frm}{\mathfrak{m}}
\newcommand{\frt}{\mathfrak{t}}
\newcommand{\CA}{\mathcal{A}}    % super gauge potential
\newcommand{\CG}{\mathcal{G}}

\newcommand{\CL}{\mathcal{L}}    % Lagrangian
\newcommand{\CF}{\mathcal{F}}

\newcommand{\CE}{\mathcal{E}}    % complex vector bundle
\newcommand{\Z}{\mathbb{Z}}    
\newcommand{\R}{\mathbb{R}}     % field of real numbers
\newcommand{\C}{\mathbb{C}}     % field of complex numbers
\newcommand{\Hbb}{\mathbb{H}}     % field of quaternions
\newcommand{\Abb}{\mathbb{A}}     
\newcommand{\CPP}{{\mathbb{C}P}}    % complex projective plane
\newcommand{\Nbb}{{\mathbb{N}}}    % complex projective plane

\newcommand{\rmF}{{\mathrm{F}}}
\newcommand{\im}{\mathrm{i}} 
 
\newcommand{\dd}{\mathrm{d}}     % total differential
\newcommand{\diag}{{\mathrm{diag}}}

\newcommand{\tr}{\mathrm{tr}}     % trace

\newcommand{\+}{\dagger}

\newcommand{\Ad}{\mathrm{Ad}} 
\newcommand{\sU}{\mathrm{U}}     
\newcommand{\sSU}{\mathrm{SU}}     
     
\newcommand{\sSO}{\mathrm{SO}}     
\newcommand{\sGL}{\mathrm{GL}}     

\newcommand{\al}{{{\alpha}}}

\newcommand{\vph}{{{\varphi}}}

\makeatletter
\renewcommand*\l@section{\@dottedtocline{1}{1.5em}{4em}}%1.5em% %2.3em%
\@addtoreset{equation}{section}
\makeatother

\begin{document}
\begin{titlepage}
\setcounter{page}{0}
.
\vskip 3cm
\begin{center}
{\LARGE \bf Stueckelberg and Higgs Mechanisms:}\\
\vskip 0.5cm
{\LARGE \bf  Frames and Scales}
\vskip 1.5cm
{\Large Alexander D. Popov}
\vskip 1cm
{\em Institut f\"{u}r Theoretische Physik,
Leibniz Universit\"{a}t Hannover\\
Appelstra{\ss}e 2, 30167 Hannover, Germany}\\
{Email: alexander.popov@itp.uni-hannover.de}
\vskip 1.1cm
\end{center}
\begin{center}
{\bf Abstract}
\end{center}
We consider Yang-Mills theory with a compact gauge group $G$ on Minkowski space $\R^{3,1}$ and compare the introduction of masses of gauge bosons using the Stueckelberg and Higgs mechanisms. The Stueckelberg field $\phi$ is identified with a $G$-frame on the gauge vector bundle $E$ and the kinetic term for $\phi$ leads to the mass of the gauge bosons. The Stueckelberg mechanism is extended to the Higgs mechanism by adding to the game a scalar field describing rescaling of metric on fibres of $E$. Thus, we associate Higgs fields as well as running coupling parameters with conformal geometry on fibres of gauge bundles. In particular, a running coupling tending to zero or to infinity is equivalent to an unbounded expansion of $G$-fibres or its contraction to a point. We also discuss scale connection, space-time dependent Higgs vacua and compactly supported gauge and quark fields as an attribute of confinement.

\end{titlepage}
\newpage
\setcounter{page}{1}

\section{Introduction and summary}

\noindent
In this paper, we address the following questions:
\begin{itemize}
\item What are the Stueckelberg fields from geometric point of view?
\item Are fundamental Higgs fields matter fields or geometry?
\item How is the Higgs mechanism different from that proposed by Stueckelberg?
\end{itemize}
It will be shown that the Stueckelberg field $\phi\in G$ defines a $G$-frame on a gauge vector bundle $E$ and the Higgs boson corresponds to a scalar field $\rho$ which defines the rescaling of frames on gauge bundles. We introduce and discuss an Abelian scale connection accompanying these rescalings. It will be shown that gauge coupling $g^{}_{\sf YM}$ is running as $\rho^{-N}$ for $G\subset\sU(N)$ with $N=1, 2, 3,...$ and fibres $G_x$ over $x\in\R^{3,1}$ of the principal gauge $G$-bundle shrink to a point for $g^{}_{\sf YM}$ growing to infinity.

We will discuss the MIT and soliton bag models~\cite{MIT, FrLee} and their generalization with space-time dependent Higgs vacua. It is proposed to relate the confinement of quarks and gluons, as well as the asymptotic freedom, to the vacuum polarization. This paper is a further development of the research program~\cite{Popov} for studying gauge and other fields defined on subspaces $S$ of space-time and vanishing outside $S$.

\section{Stueckelberg mechanism}

\noindent
{\bf Vector bundle $E$.}  We consider Minkowski space $M=\R^{3,1}$ with the metric
\begin{equation}\label{2.1}
\dd s^2_M=\eta^{}_{\mu\nu}\dd x^\mu\dd x^\nu,\quad \eta =(\eta^{}_{\mu\nu})=\diag (-1, 1, 1, 1)\ ,
\end{equation}
where $x^\mu$ are coordinates on $M$, $\mu ,\nu = 0,...,3$. Let $G$ be a compact Lie group, $\frg$ its Lie algebra and $P(M,G)=M\times G$ a trivial principal $G$-bundle over $M$. We introduce a complex vector bundle $E=P\times_G V$, where $V=\C^N$ is the space of irreducible representation of $G$, and endow the space $V$ with a Hermitian scalar product
\begin{equation}\label{2.2}
\langle \psi , \psi\rangle=\psi^\+\psi =\delta_{i\bar\jmath}\psi^i\bar\psi^{\bar\jmath}\ ,
\end{equation}
where $\psi\in\C^N$ are sections of $E$. Thus, $E$ is a Hermitian vector bundle and $G$ can be considered as a closed subgroup of $\sU(N)$. For $N=1$ we consider $G=\sU(1)$ and for $N>1$ we have in mind the group $G=\sSU(N)$. We choose the normalization of the generators $I_a$ of the group $G$ such that $\tr (I_aI_b)=-\delta_{ab}$ for any representation, $a,b=1,...,\dim G$.

\smallskip

\noindent
{\bf Connections and automorphisms.} Let $\CA =\CA_\mu\dd x^\mu$ be a connection one-form (gauge potential) on $E$, and $\CF = \dd\CA + \CA\wedge\CA =\sfrac12\,\CF_{\mu\nu}\dd x^\mu\wedge\dd x^\nu$ its curvature (gauge field) taking values in the Lie algebra $\frg =\,$Lie$\,G$. On each fibre $E_x\cong\C^N$ of the bundle $E\to M$ the group $G_x$ acts by rotations of the basis in $E_x$, $x\in M$. There is a one-to-one correspondence between the group $G_x$ and all {\it ordered basis}, or {\it frames}, on the fibres $E_x$. Thus, $G$-frames on $E$ are parametrized by the infinite-dimensional group
\begin{equation}\label{2.3}
\CG =C^\infty (M,G)
\end{equation}
of smooth $G$-valued functions on $M=\R^{3,1}$. This is the group of base-preserving automorphisms Aut$_GE$ of the bundle $E\to M$ and its Lie algebra is Lie$\,\CG = C^\infty (M, \frg )$.

 \smallskip

\noindent
{\bf Remark.} One should not confuse frames and automorphisms of the bundle $P(M, G)$, even if they are indistinguishable  in the case of a trivial bundle $P(M,G)=M\times G$. For a curved manifold $M$, the frame bundle $P(M,G)$ has only local sections. At the same time, the bundle of groups Inn$\,P=P\times_GG$, where $G$ acts on itself by inner automorphisms $f\mapsto gfg^{-1}$ for $f,g\in G$, has global sections -- they are automorphisms of the bundle $P(M,G)$. For trivial bundles both of these spaces are parametrized by the group \eqref{2.3}, but their geometric meaning is different. 

We denote by $\Abb$ the space of all smooth connections on $E$.  The group $\CG$ of $P$-automorphisms acts on $\CA\in\Abb$ by the standard formula
\begin{equation}\label{2.4}
\CA\mapsto\CA^g=g^{-1}\CA g+ g^{-1}\dd g 
\end{equation}
for $g\in \CG$ and $\dd = \dd x^\mu\partial /\partial x^\mu$. Whether $\CG$ (or its subgroup) is the group of gauge transformations or a dynamical group depends on the choice of Lagrangian and boundary conditions~\cite{Popov}.

\smallskip

\noindent
{\bf Lagrangian.} Let us consider the Lagrangian density for massive gauge field,
\begin{equation}\label{2.5}
\CL^{}_{\sf YM}+\CL^{}_{m}=\frac{1}{4g^2_*}\,\tr (\CF_{\mu\nu}\CF^{\mu\nu})+
\frac{1}{2}\,v^2\eta^{\mu\nu}\tr (\CA_{\mu}\CA_{\nu})\ ,
\end{equation}
where $g^{}_*:=g^{}_{\sf YM}>0$ is the gauge coupling constant and $m=v g^{}_*\ge 0$ is the mass parameter. Space-time indices here and everywhere are raised using the metric \eqref{2.1}. The massless case corresponds to $v=0$.

The mass term $\CL^{}_{m}$ in \eqref{2.5} with $v\ne 0$ explicitly breaks the invariance of the Lagrangian  \eqref{2.5} under the transformation  \eqref{2.4} from the group  \eqref{2.3} of $G$-automorphisms. The Lagrangian density  \eqref{2.5} describes massive gauge bosons having two transverse components $\CA^T$ and one longitudinal component $\CA^L$ of gauge potential\footnote{The component $\CA_0$ is nondynamical.} $\CA = \CA_\mu\dd x^\mu$. In this case the group \eqref{2.3} acts on $\CA$ as a dynamical group, i.e. it maps $\CA\in\Abb$ to $\CA^g\in\Abb$ which is not equivalent to $\CA$. 

\smallskip

\noindent
{\bf Stueckelberg field.} The Stueckelberg field $\phi$ is a {\it frame} on $E$, i.e. a $G$-valued function on $M$ which is parametrized by elements from the group \eqref{2.3}. For example, for $\C^N$-bundle $E\to M$ and $G=\sSU(N)$, the frame is given by $N$ basis vectors $\phi_i\in\C^N$ such that
\begin{equation}\label{2.6}
\phi_1^\+\phi_1=...=\phi_N^\+\phi_N=\unit_N,\quad \phi_i^\+\phi_j=0\ \ \mbox{for}\ \ i\ne j\quad \Rightarrow\ \  \phi =(\phi_1...\phi_N)\in \sSU(N). 
\end{equation}
The Stueckelberg fields can be pointwise multiplied as
\begin{equation}\label{2.7}
\phi\mapsto\phi^g:=g^{-1}\phi\in\CG\ \ \mbox{for}\ \ g, \phi\in\CG\ . 
\end{equation}
This is the right action of $\CG$ on itself. Thus, these fields are elements of the group $\CG$ which is a dynamical group in the case of nonzero mass of gauge bosons.

Let us map $\CA\in\Abb$ into $\CA^\phi$ for $\phi\in\CG$ as
\begin{equation}\label{2.8}
\CA\mapsto\CA^\phi =\phi^{-1}\CA\phi + \phi^{-1}\dd\phi
\end{equation}
and consider the term
\begin{equation}\label{2.9}
\CL_\phi = \sfrac12\, v^2\eta^{\mu\nu}\tr (\CA_\mu^\phi\CA_\nu^\phi )=  \sfrac12\, v^2\tr (\CA_\mu -\phi\partial_\mu\phi^\+)(\CA^\mu - \phi\partial^\mu\phi^\+ )=-\sfrac12\, v^2\tr (\nabla_\mu\phi)^\+\nabla^\mu\phi \ ,
\end{equation}
where $\phi^\+ =\phi^{-1}$ and
\begin{equation}\label{2.10}
\nabla_\mu\phi := \partial_\mu\phi +\CA_\mu\phi\ .
\end{equation}
It is easy to see that under the action of $\CG$ on $(\CA ,\phi)$ given by \eqref{2.4} and \eqref{2.7}, we have
\begin{equation}\label{2.11}
\CA^\phi \mapsto (\CA^g)^{\phi^g} =\CA^\phi =:\CA^{\sf inv}
\end{equation}
and
\begin{equation}\label{2.12}
\CA^{\sf adj}:=\CA - \phi\dd\phi^\+\ \mapsto\ \CA^g - \phi^g\dd(\phi^g)^\+= g^\+\CA^{\sf adj}g\ .
\end{equation}
Thus, the dressed\footnote{The map \eqref{2.8} is the {\it dressing transformation} considered e.g. in~\cite{McM, Franc}. An overview of the dressing field method in gauge theories and many references can be found in~\cite{Berg}.} gauge potential $\CA^\phi$ is invariant under automorphism group $\CG$.

\smallskip

\noindent
{\bf Mass term.} For both \eqref{2.11} and \eqref{2.12} the mass term \eqref{2.9} is invariant under the action of the group $\CG$. This term reduces to the term $\CL_m$ in \eqref{2.5} after transforming $\phi\to\unit_N$ (gauge fixing). Hence, we can consider the $\CG$-invariant theory with the Lagrangian density
\begin{equation}\label{2.13}
\CL^{}_{\sf YM}+\CL^{}_{\phi}=\frac{1}{4g^2_*}\,\tr \left\{\CF_{\mu\nu}\CF^{\mu\nu}-
2m^2 (\nabla_{\mu}\phi)^\+\nabla^{\mu}\phi\right\}\ ,
\end{equation}
with the Stueckelberg field $\phi\in G$. In other words, we can equally consider either $\CG$-invariant model \eqref{2.13} with $(\CA^T, \CA^L=0, \phi)$ or its gauge-fixed version \eqref{2.5} with $(\CA^T, \CA^L\ne 0, \phi=\unit_N)$, i.e. $\phi$ parametrizes the longitudinal components $\CA^L$ of gauge potentials $\CA$. For the Abelian case $G=\sU(1)$, this trading of degrees of freedom was proposed by Stueckelberg \cite{Stueck} in 1938 and then, 25 years later, it was rediscovered as the Goldstone fields and the Higgs mechanism \cite{BEH} (for a historical overview and references see e.g. \cite{Berg, Ruegg}).

\smallskip

\noindent
{\bf Remark.} Note that \eqref{2.13} is a Higgs-type Lagrangian with a $G$-valued field $\phi$. Some sources claim that Stueckelberg proposed an ``affine Higgs mechanism'' in which the compact group $G=\sU(1)$ is replaced by the non-compact group $\sGL^+(1, \R)=\R^+$. This is nonsense, he introduced the field $\phi\in G$ which is now called the Nambu-Goldstone boson.

\smallskip

\noindent
{\bf Framed bundles.} As we discussed, $\phi$ can be identified with an element of the group $\CG=\mbox{Aut}_GE$ of automorphisms of the vector bundle $E$. The group $\CG$ rotates frames on $E$. Recall that \eqref{2.5} is a gauge fixed version of \eqref{2.13}, where $\CG$-invariance is unbroken. Hence, we can consider a {\it fixed} bundle $E_0$ associated with \eqref{2.5} and an unfixed bundle $E$ associated with \eqref{2.13}.  Then $\phi$ defines an {\it isomorphism} of these bundles,
\begin{equation}\label{2.14}
\phi : \ \  E\ \to\ E_0\ .
\end{equation}
The bundle $E$ with the isomorphism \eqref{2.14} is called a {\it framed bundle}. This equivalent description of Stueckelberg fields can be used  when considering them at boundaries ({\it edge modes} \cite{GSW, DF, Str}) and has been discussed in detail in \cite{Popov}.

\smallskip

\noindent
{\bf Unbroken subgroup $H\subset G$.} Let $\{I_a\}$  with $a=1,...,\dim G$ be the generators of the Lie group $G$, normalized such that $\tr (I_aI_b)=-\delta_{ab}$. We can expand $\CA = \CA^aI_a$ in terms of the Lie algebra $\frg$ basis $\{I_a\}$. Above, we described the Stueckelberg mechanism for the {\it complete breaking} of gauge invariance and the acquisition of mass by all gauge bosons $\CA^a$, $a=1,...,\dim G$. Below we will consider how to keep massless bosons from a closed subgroup $H$ of the group $G$ and introduce masses for only the $G/H$-part of all bosons.

\smallskip

\noindent
{\bf Orbits $G/H$}. Consider a closed subgroup $H$ of $G{\subset}\sU(N)$ such that $G/H$ is a reductive homogeneous space. The Lie algebra $\frg$ of $G$ can be decomposed as $\frg =\frh\oplus\frm$, where $\frm$ is the orthogonal  complement of the Lie algebra $\frh =\,$Lie$H$ in $\frg$. We denote by $\frt$ the Cartan subalgebra of $\frg$ and by $\frt_+\subset\frt$ the positive closed Weyl chamber. Let us choose an element $\xi_0\in\frt_+$ such that
\begin{equation}\label{2.15}
h\xi_0h^\+=\xi_0
\end{equation}
for $h\in H$, i.e. $H$ is a stabilizer of $\xi_0$. Then the adjoint orbit\footnote{We identify the algebra $\frg$ and the space $\frg^*$ dual to $\frg$ via a scalar product tr on $\frg$ and consider adjoint orbits instead of coadjoint ones.}
\begin{equation}\label{2.16}
O_{\xi_0}=\bigl\{\vph=g\xi_0g^\+=\phi\xi_0\phi^\+\in\frg\ \mid\  g=\phi h\in G,\ \phi\in G/H\bigr\}
\end{equation}
will be diffeomorphic to the coset space $G/H$. The definition   \eqref{2.16} gives a parametrization of one of the patches covering the orbit $O_{\xi_0}\cong G/H$. One can extend parametrization to all other patches by the action of the Weyl group of $G$.

\smallskip

\noindent
{\bf Flag manifolds.} A special case of orbits \eqref{2.16} is related with a flag structure in the complex vector space $V=\C^N$. A flag of $V$ is a filtration $V_1\subset V_2\subset ... \subset V_k$, where $V_i=\C^{d_i}$, $0\le d_1 < d_2 < ... < d_k$ and $k\le N$. We will not discuss a manifold of flags in $V$ in full generality (parabolic subgroups of $G^{\C}$, highest weight representations etc.). We consider only the space $\C^N$ with a Hermitian metric preserved by the group $G=\sU(N)$. In this case one should define the splitting
\begin{equation}\label{2.17}
\C^N=\C^{N_1}\oplus ... \oplus\C^{N_k},\  \  N_1+...+N_k=N,
\end{equation}
preserved by the subgroup $H=\sU(N_1)\times ... \times \sU(N_k)\subset \sU(N)$ and then the flag manifold will be
\begin{equation}\label{2.18}
O_{\xi_0}=\sU(N)/\sU(N_1)\times ... \times \sU(N_k)
\end{equation}
with $\xi_0=\im (\al_1\unit_{N_1}, ... , \al_k\unit_{N_k})$ and $N_i=d_i-d_{i-1}, d_0=0$. For example, with $\C^2$ and $\C^3$ one can associate the following homogeneous spaces:
\begin{equation}\label{2.19}
\C^2=\C\oplus\C\ \Rightarrow \ O_{\xi_0}=\sU(2)/\sU(1)\times \sU(1)=\sSU(2)/\sU(1)\cong\CPP^1,
\end{equation}
\begin{equation}\label{2.20}
\C^3=\C\oplus\C^2\ \Rightarrow \ O_{\xi_0}=\sU(3)/\sU(1)\times \sU(2)=\sSU(3)/\sU(2)\cong\CPP^2,
\end{equation}
\begin{equation}\label{2.21}
\C^3=\C\oplus\C\oplus\C\ \Rightarrow \ O_{\xi_0}=\sU(3)/\sU(1)\times \sU(1)\times \sU(1)=\sSU(3)/\sU(1)\times \sU(1)\ .
\end{equation}
All of them are K\"ahler manifolds.

\smallskip

\noindent
{\bf Mass terms for $\CA\in\frm\subset\frg$.} If we want to introduce masses only for gauge bosons parametrized by the coset space $G/H$, we can start from the mass term
\begin{equation}\label{2.22}
\CL_{\frm}=\sfrac12\eta^{\mu\nu}\tr [\CA^0_\mu , \xi_0][\CA^0_\nu , \xi_0]
\end{equation}
which is invariant under transformations from the subgroup $C^\infty (M,H)$ of the group $\CG = C^\infty (M,G)$. Generating the orbit \eqref{2.16} by dressing formula
\begin{equation}\label{2.23}
\vph_0=\xi_0\ \mapsto\ \vph=g\xi_0 g^\+\quad\mbox{and}\quad\CA^0\ \mapsto\ \CA = g\CA^0g^\++g\dd g^\+\ ,
\end{equation}
we transform \eqref{2.22} to the $\CG$-invariant mass term
\begin{equation}\label{2.24}
\CL_{\vph}=\sfrac12\,\tr\nabla_\mu\vph\nabla^\mu\vph\quad\mbox{with}\quad\nabla_\mu\vph :=\partial_\mu\vph + [\CA_\mu, \vph ]\ .
\end{equation}
Here the $\frg$-valued field $\vph$ is parametrized by the Stueckelberg field $\phi$ with values in the coset space $G/H$ according to \eqref{2.16}. Again we obtain a Higgs-type Lagrangian density
\begin{equation}\label{2.25}
\CL^{}_{\sf YM}+\CL^{}_{\vph}=\frac{1}{4g^2_*}\,\tr \left\{\CF_{\mu\nu}\CF^{\mu\nu}+
2g_*^2 \nabla_{\mu}\vph\nabla^{\mu}\vph\right\}\ ,
\end{equation}
but now we get masses only for gauge bosons $\CA\in\frm=\mbox{Lie}(G/H)$. The mass term \eqref{2.22} is the gauge fixed form of \eqref{2.24}.

Note that if we consider $G=\sU(N)$ and the Hermitian vector bundle $E$ then the subgroup $H$ of $G$ preserves the flag structure \eqref{2.17} in fibres $E_x\cong\C^N$. The Stueckelberg field $\phi\in G/H$ parametrizes this flag structure \eqref{2.18}.

\section{Higgs mechanism}

\noindent
Lagrangians \eqref{2.13} and \eqref{2.25} are standard for gauge fields interacting with $G$-valued scalar field $\phi$ or $\frg$-valued field $\vph=\phi\xi_0\phi^\+$. In both cases $\phi^\+\phi =\unit_N$ and fixing the gauge $\phi=\unit_N$ one can get mass terms \eqref{2.5} and \eqref{2.22}, respectively. From now on, we will consider the case \eqref{2.13} with $\phi\in G$ and return to the case \eqref{2.25} later.

\smallskip

\noindent
{\bf Scaling field $\rho$.} As discussed in Section 2, the Stueckelberg fields $\phi$ can be identified with automorphisms of the gauge bundle $E, \phi\in\mbox{Aut}_GE$. In other words, we can consider $\phi$ as a map from a fixed frame to the new ones, i.e. as a rotation of bases on fibres $E_x\cong\C^N$ of $E$. These rotations act on sections $\psi$ of this bundle as mapping
\begin{equation}\label{3.1}
\phi :\ \psi\ \mapsto\ \psi^\phi = \phi\psi\ .
\end{equation}
Recall that the bundle $E$ is Hermitian, with the metric \eqref{2.2} on fibres. Obviously, this metric is invariant under rotations \eqref{3.1}.

Let us now consider a rescaling of metric \eqref{2.2} as a mapping
\begin{equation}\label{3.2}
h(\psi)=\langle\psi , \psi\rangle\ \mapsto\ \widetilde h=\rho^2h=\rho^2\langle\psi , \psi\rangle = \langle\rho\psi , \rho\psi\rangle=\langle\widetilde\psi , \widetilde\psi\rangle\ ,
\end{equation}
where $\rho (x)>0$ is a function of $x\in M$. The maps \eqref{3.1} and \eqref{3.2} can be combined into the map
\begin{equation}\label{3.3}
\Phi = \rho\phi :\ \ \psi\mapsto\widetilde\psi = \Phi\psi\ \ \mbox{with}\ \ \rho\in\sGL^+(1,\R)=\R^+\ \mbox{and}\ \phi\in G .
\end{equation}
Matrix $\Phi$ in \eqref{3.3} is an element of the conformal extension of the group $G{\subset}\sU(N)$ defined \cite{Kob} as
\begin{equation}\label{3.4}
\widetilde G=\R^+\times G=\bigl\{\Phi = \rho\phi , \ \rho\in\R^+, \phi\in G \mid \Phi^\+\Phi =\rho^2\unit_N\bigr\}\ .
\end{equation}
As a manifold, the group \eqref{3.4} is a cone $C(G)$ over $G$ with the metric
\begin{equation}\label{3.5}
\dd s^2_{\widetilde G}=\dd\rho^2 + \rho^2\dd s^2_G\ .
\end{equation}
If we add the tip $\Phi =0$ to the cone $C(G)$ then we get a {\it semigroup} with identity (a monoid), since the element $\Phi=0$ has no inverse.

\smallskip

\noindent
{\bf Special point $\rho =0$.} The function $\rho$ in \eqref{3.2}-\eqref{3.5} defines a {\it scale} on fibres $E_x\cong\C^N$ of the Hermitian vector bundle $E\to M$ and $\widetilde G$ is a {\it conformal structure} on fibres of $E$. Note that the cone $C(G)$ can be projected onto $\R^+$,
\begin{equation}\label{3.6}
\pi :\  C(G)\stackrel{G}{\longrightarrow}\ \R^+=C(G)/G\ \mbox{by}\ \Phi\to\rho=(\Phi^\+\Phi)^{1/2}
\end{equation}
with fibres $G_\rho$ over $\rho\in\R^+$ since $C(G)$ is a cohomogeneity one Riemannian $G$-manifold. The orbit $G_\rho$ for $\rho\to 0$ is singular that is obvious from \eqref{3.5}. At this point both $G_\rho$ and fibres $\C^N$ of $E$ shrink to a point. 

\smallskip

\noindent
{\bf Examples.} For $G=\sU(1)$ we have $\widetilde G=\R^+\times\sU(1)\cong\C^*$ which is the multiplicative group of non-zero complex numbers. Adding the point $\rho =0$ corresponds to the transition from $\C^*$ to the field $\C=\C^*\cup\{0\}$ of all complex numbers. For $G=\sSU(2)$, the conformal extensions give the group of non-zero quaternions $\Hbb^*=C(S^3)\cong\R^4{\setminus}\{0\}\cong\C^2{\setminus}\{0\}$ and adding the point $\rho=0$ we get the semigroup $\Hbb\cong\C^2\cong\R^4$.

\smallskip

\noindent
{\bf Mass term with $\rho$.} Recall that massive gauge bosons are described by the Lagrangian \eqref{2.5}. The longitudinal components of these bosons can be transferred to the Stueckelberg field $\phi$ to obtain the gauge-invariant mass term \eqref{2.9}. To see the effect of {\it rescaling} \eqref{3.2} of the metric on fibres $E_x\cong\C^N$ of the bundle $E$, we replace $\phi\in G$ in \eqref{2.9} with $\Phi=\rho\phi\in\widetilde G$. Then we obtain the term
\begin{equation}\label{3.7}
-\sfrac12\,\tr (\nabla_\mu\Phi)^\+\nabla^\mu\Phi = -\sfrac12\,\rho^2\tr (\nabla_\mu\phi)^\+\nabla^\mu\phi -\sfrac12\,N\partial_\mu\rho\partial^\mu\rho
\end{equation}
coinciding with  \eqref{2.9} if we put $\rho =v=\,$const. Thus, the scaling field $\rho$ becomes dynamical, and we can add to \eqref{3.7} a potential term (self-action) of the form\footnote{The fields $\phi$ and $\rho$, as well as the kinetic terms for them, are introduced in a natural way, but the choice of potential energy $V(\rho )$ is quite arbitrary. This is a drawback.}
\begin{equation}\label{3.8}
V(\rho )=-\frac{\lambda}{4}\Bigl (\frac{1}{N}\,\tr(\Phi^\+\Phi ) - v^2\Bigr)^2=-\frac{\lambda}{4}\bigl (\rho^2 - v^2\bigr)^2\ .
\end{equation}
Putting together the resulting terms, we obtain the Lagrangian density
\begin{equation}\label{3.9}
\CL = \frac{1}{4g_*^2}\tr(\CF_{\mu\nu}\CF^{\mu\nu}) - \frac{1}{2}\,\rho^2\tr (\nabla_\mu\phi)^\+\nabla^\mu\phi - \frac{N}{2}\partial_\mu\rho\partial^\mu\rho - \frac{\lambda}{4}\,(\rho^2 - v^2)^2
\end{equation}
for Yang-Mills-Higgs theory. From \eqref{3.9} it follows that for $N=2$ we can identify the scaling function $\rho$ with the Higgs boson.

\smallskip

\noindent
{\bf Higgs field in SM.} The Standard Model (SM) uses the Higgs field $\psi$ with values in the fundamental representation $\C^2$ of the group SU(2). We will show that this is equivalent to the description \eqref{3.2}-\eqref{3.9} for $G=\sSU(2)\cong S^3$ and $\widetilde G=C(G)=\Hbb^*\cong\C^2{\setminus}\{0\}$.\footnote{We consider point $\{0\}$ as the limit $\rho\to 0$.}

Let us introduce the matrix
\begin{equation}\label{3.10}
\Phi =\rho\phi = \rho\begin{pmatrix}\bar b&a\\-\bar a&b\end{pmatrix} =: (\hat\psi \psi )
\end{equation}
with $\phi\in\sSU(2)$. We have
\begin{equation}\label{3.11}
\hat\psi = \Phi\begin{pmatrix}1\\0\end{pmatrix},\ \psi = \Phi\begin{pmatrix}0\\1\end{pmatrix},\ \Phi^\+\Phi=\rho^2\unit_2\ \Leftrightarrow\ \hat\psi^\+\hat\psi = \rho^2=\psi^\+\psi ,\ \hat\psi^\+\psi =0
\end{equation}
and therefore\footnote{Note that dressing transformations for $\psi$ from \eqref{3.10} were considered in \cite{Berg}.}
\begin{equation}\label{3.12}
\tr(\nabla_\mu\Phi)^\+\nabla^\mu\Phi = 2(\nabla_\mu\hat\psi)^\+\nabla^\mu\hat\psi= 2(\nabla_\mu\psi)^\+\nabla^\mu\psi\ .
\end{equation}
The last term in \eqref{3.12} gives the kinetic term for $\psi\in\C^2$ and from  \eqref{3.11} we see that $\psi^\+\psi =\rho^2$. In SM the vacuum state is usually chosen as $a=0$, $b=1$ and $\rho =v$, so that
\begin{equation}\label{3.13}
\psi=\Phi\begin{pmatrix}0\\1\end{pmatrix}=\begin{pmatrix}0\\v\end{pmatrix}\ \ \Rightarrow\ \ \Phi =v\unit_2\ .
\end{equation}
If we want to keep in \eqref{3.12} only $\psi$ then we can use in \eqref{3.12} the term
\begin{equation}\label{3.14}
\tr(\nabla_\mu\Phi)^\+(\nabla^\mu\Phi)P\quad\mbox{with}\quad  P=\begin{pmatrix}0&0\\0&1\end{pmatrix}\ ,
\end{equation}
where $P$ is a projector. The representation of the Higgs field in the form \eqref{3.10} is important for a better understanding of the mechanism for generating a mass of gauge bosons. From \eqref{3.10} we see that $\Phi$ defines a conformal frame on $\C^2$-bundle $E$ and $\rho$ sets a scale on fibres of $E$. 

\smallskip

\noindent
{\bf Remark.} Suppose we have two vector bundles, $E_1$ of rank $N_1$ and $E_2$ of rank $N_2$, associated with a principal $G_1$-bundle and a $G_2$-bundle, respectively. Then we naturally have two scaling bosons, $\rho_1$ and $\rho_2$, since they are related to the geometry of bundles, and are not introduced artificially. However, one or both of them can be ``frozen'' to a constant value. 

\smallskip

\noindent
{\bf Algebra-valued Higgs fields.} In the case of masses for a part $\CA\in\frm\subset \frg$ of gauge bosons, discussed in \eqref{2.15}-\eqref{2.25}, one can generalize \eqref{2.24} as follows. Replace $\xi_0$ by a function $\xi (x)\in \frt_+\subset\frt\subset\frg$. Instead of 
\eqref{2.22} one should start with the term
\begin{equation}\label{3.15}
\sfrac12\,\eta^{\mu\nu}\tr(\partial_\mu\xi + [\CA_\mu^0, \xi])(\partial_\nu\xi + [\CA_\nu^0, \xi])
\end{equation}
and dress $\xi$ via \eqref{2.16}. After this we again obtain \eqref{2.24} and \eqref{2.25}, but with $\vph$ parametrized not only by $\phi\in G/H$ but also the $\frt$-valued function $\xi$. A potential term for such $\vph$ can be introduced in term of $\xi=\xi^iI_i$, e.g. as
\begin{equation}\label{3.16}
V(\xi )=\sfrac14\, \sum\limits_{i=1}^{{\sf rank} G}\lambda_i\bigl ((\xi^i)^2 - v^2_i\bigr)^2\ .
\end{equation}
For cosets of type \eqref{2.18} one can take $\xi = \im \bigl(\xi_1(x)\unit_{N_1}, ... , \xi_k(x)\unit_{N_k}\bigr)$ with $\tr\,\xi =0$. The fields $\xi_1,...,\xi_k$ define non-uniform rescaling on subspaces $\C^{N_1},...,\C^{N_k}$ of $\C^N$.

Summing up preliminary results, we can say that
\begin{itemize}
\item $G$-valued Stueckelberg fields $\phi$ parametrize {\it frames} on gauge vector bundles $E$ over $M$,
\item fields $\rho$ in $\Phi =\rho\phi\in\widetilde G$ define {\it scales} on fibres $E_x\cong\C^N$ of $E$,
\item in massless gauge theory, fields $\phi$ define gauge transformations,
\item  in massive gauge theory, fields $\phi$ define longitudinal components of gauge potentials,
\item vacuum value of the {\it scale field} $\rho$ sets the mass of gauge bosons. 
\end{itemize} 
The limit $\rho\to 0$ in  \eqref{3.9} nullifies the mass term.

\section{Rescaling and scale gauge fields}

\noindent
{\bf Conformal geometry.} Let $M$ be an $m$-dimensional smooth Riemannian or Lorentzian manifold with a metric $g$. Consider a smooth positive function $\Omega$ on $M$ and define the metric
\begin{equation}\label{4.1}
\widetilde g = \Omega^2g\ .
\end{equation}
The metric $\widetilde g$ is called conformally equivalent to the metric $g$ and an equivalence class $[g]=\{\widetilde g\sim g\}$ of such metrics is called a {\it conformal structure},
\begin{equation}\label{4.2}
[g]= \{\Omega^2g \mid \Omega(x)>0\}\ .
\end{equation}
Considering rescaling \eqref{4.1}, Hermann Weyl introduced a conformal generalization of Riemannian geometry.\footnote{For a historical overview and references see e.g. \cite{Scholz}.} A conformal structure on manifolds with $m>3$ is locally flat if the Weyl curvature tensor of some (and hence any) Riemannian metric $g$ from a class $[g]$ is zero. It is globally conformally flat if also the Riemannian curvature tensor vanishes.

By considering rescaling \eqref{4.1}, Weyl introduced the generalized Christoffel symbols
\begin{equation}\label{4.3}
\widetilde\Gamma^\sigma_{\mu\nu}=\Gamma^\sigma_{\mu\nu}+\delta^\sigma_\mu w_\nu + \delta^\sigma_\nu w_\mu -g_{\mu\nu}g^{\sigma\lambda}w_\lambda\ ,
\end{equation}
where $\mu , \nu ,... = 0, ... , m-1$, $\Gamma^\sigma_{\mu\nu}$ are standard Christoffel symbols and $w=w_\mu \dd x^\mu$ is a Weyl connection transforming when rescaling \eqref{4.1} by the formula
\begin{equation}\label{4.4}
w_\mu \mapsto \widetilde w_\mu = w_\mu - \partial_\mu \ln\Omega\ .
\end{equation}
Using this Abelian connection, one can introduce a scale covariant derivative on $M$. The above conformal generalization of (pseudo-)Riemannian geometry is being actively developed both in mathematics and in physics.

\smallskip

\noindent
{\bf Scaling geometry.} In Section 3, we showed that the Higgs boson field $\rho$ defines the rescaling \eqref{3.2} of $\C^N$-vectors $\psi$ and the metric $h$ on fibres $E_x\cong\C^N$ of the gauge vector bundle $E$. The scale parameter $\rho$ depends on $x\in M=\R^{3,1}$ and hence it is reasonable to introduce a scale connection
\begin{equation}\label{4.5}
a=a_\mu\dd x^\mu\ \ \Rightarrow\ \ f:=\dd a =\sfrac12\,f_{\mu\nu}\dd x^\mu\wedge\dd x^\nu =\sfrac12\,(\partial_\mu a_\nu - \partial_\nu a_\mu)\dd x^\mu\wedge\dd x^\nu
\end{equation}
similar to the Weyl connection $w=w_\mu\dd x^\mu$ in \eqref{4.4}. The Abelian connection $a$ is defined on the principal bundle $P(M, \sGL^+(1, \R))$ with the multiplicative group $\sGL^+(1, \R))=\R^+$ and on the associated real line bundle $L_+=M\times\R^+$.

In contrast to the Weyl connection $w$, the connection $a$ is associated not with rescaling the metric on space $T_xM$ tangent to space-time $M$, but with rescaling the metric on internal vector spaces $E_x$. We consider the tensor product $\widetilde E=E\otimes L_+$ of bundles $E$ and $L_+$ associated with principal bundle $\widetilde P (M, \widetilde G)=M\times\widetilde G$ for $\widetilde G =\R^+\times G$. One-form of connection on this bundle is
\begin{equation}\label{4.6}
\widetilde \CA = \CA\otimes 1 + \unit_N\otimes a
\end{equation}
and for simplicity we will write it as $\widetilde \CA = \CA +a$.  The covariant derivative $\widetilde\nabla$ of sections $\widetilde\psi$ of $\widetilde E$ reads as\footnote{Note that one can also consider bundles $L_+^\ell$, sections $\zeta$ of which have {\it scale weight} $\ell$ with covariant derivative $D_\mu\zeta =\partial_\mu\zeta + \ell a_\mu\zeta$.}
\begin{equation}\label{4.7}
\widetilde\nabla_\mu\widetilde\psi = \partial_\mu\widetilde\psi + (\CA_\mu + a_\mu )\widetilde\psi  =  \nabla_\mu\widetilde\psi + a_\mu\widetilde\psi\ .
\end{equation}
When rescaling sections of $\widetilde E$, we have
\begin{equation}\label{4.8}
\widetilde\psi \mapsto \widetilde\psi^\prime =\rho\widetilde\psi \quad\mbox{and}\quad a\mapsto a^\prime =a - \dd\ln\rho \ ,
\end{equation}
with $\widetilde\nabla^\prime\widetilde\psi^\prime = \rho\widetilde\nabla\widetilde\psi$.

The field $\Phi$ in \eqref{3.3}-\eqref{3.8} has scale weight one and hence the covariant derivative of the form \eqref{4.7}. Substituting $\widetilde\nabla\Phi$ into \eqref{3.7}, we obtain the same formula with replacement
\begin{equation}\label{4.9}
\partial_\mu\rho\ \mapsto\  D_\mu\rho = \partial_\mu\rho + a_\mu\rho = \rho (a_\mu + \partial_\mu\ln\rho ).
\end{equation}
Substituting this into \eqref{3.9} and adding the standard Lagrangian for the Abelian gauge field $f=\dd a$, we obtain the Lagrangian density
\begin{equation}\label{4.10}
\CL =\frac{1}{4g_*^2}\,\tr (\CF_{\mu\nu}\CF^{\mu\nu}) - \frac{1}{4}\,f_{\mu\nu}f^{\mu\nu}- \frac{1}{2}\,\rho^2\tr (\nabla_\mu\phi)^\+\nabla^\mu\phi - \frac{N}{2}\,D_\mu\rho D^\mu\rho - \frac{\lambda}{4}\,(\rho^2-\rho_0^2)^2\ ,
\end{equation}
where $\rho_0(x)$ is a fixed function.

Vacua of the model \eqref{4.10} are given by flat connections
\begin{equation}\label{4.11}
\CA = \phi\dd\phi^\+ \quad\mbox{and}\quad a= - \dd\ln\rho 
\end{equation}
at the minimum $\rho =\rho_0$ of the potential $V(\rho )$. Hence, the field
\begin{equation}\label{4.12}
\Phi_0=\rho_0\phi
\end{equation}
is an arbitrary $\widetilde G$-valued  function defining the vacuum bundle $E_0$ with the flat connections \eqref{4.11} for which $\CF=0=f$. The standard vacuum arises as a special case when choosing $\rho_0=v=\,$const$\ \ \Rightarrow\ a=0$. The possibility of vacuum states parametrized by coordinate-dependent scaling function $\rho =\rho_0$ appears due to the introduction of new degrees of freedom given by the scale connection \eqref{4.5}.

\smallskip

\noindent
{\bf Remark.}  Let us introduce a dressed scale connection
\begin{equation}\label{4.13}
a^\rho = a+\dd\ln\rho\ .
\end{equation}
Then from \eqref{4.9} we obtain
\begin{equation}\label{4.14}
D_\mu\rho = \rho\, a_\mu^\rho\quad\mbox{and}\quad f_{\mu\nu}^\rho = f_{\mu\nu}
\end{equation}
so that \eqref{4.10} can be rewritten in terms of $a_\mu^\rho$ and $f_{\mu\nu}^\rho$ which are invariant under the transformations
\begin{equation}\label{4.15}
\rho\ \mapsto\ \widetilde\rho =\gamma\rho
\end{equation}
for real-valued functions $\gamma\in C^\infty (\R^{3,1}, \R^+)$.  Then there will be no derivatives of $\rho$ in  \eqref{4.10}. It is not yet clear which description is better to use. From  \eqref{4.14} one can always return to the standard formulation with $\partial_\mu\rho\,\partial^\mu\rho$ by considering the flat connection $a^\rho =\dd\ln\rho$. 

Under scaling transformations \eqref{4.15} the fields $A, \phi$ are not transformed, but due to the factor $\rho^2$ in several terms in \eqref{4.10}, the Lagrangian is not invariant under  the transformations \eqref{4.15}. Note that the space-time scale transformations \eqref{4.1}-\eqref{4.4} and scale transformations \eqref{3.2}-\eqref{3.4}, \eqref{4.8} and \eqref{4.15} of internal spaces are independent and all fields can have different scale weight with respect to these scaling groups.

\section{Running couplings and adjoint bundle}

\noindent
{\bf Group manifolds.} Let $\{I_a\}$ with $a=1,...,\dim G$ be the generators of the Lie group $G$ with structure constants
$f^c_{ab}$ given by the commutation relations
\begin{equation}\label{5.1}
[I_a,I_b]=f^c_{ab}I_c\ .
\end{equation}
We can normalize $I_a$ such that the Killing-Cartan metric on $\frg =\,$Lie$\,G$ is $f_{ad}^cf^d_{cb} =\delta_{ab}$.

For group elements $g\in G$ not depending on $x\in\R^{3,1}$ we introduce left- and right-invariant one-forms on $G$,
\begin{equation}\label{5.2}
g^{-1}\dd_Gg=:\theta^a_LI_a\quad\mbox{and}\quad (d_Gg)g^{-1}=:\theta^a_RI_a\ ,
\end{equation}
where $\dd_G$ is the exterior derivative on $G$. Then for the metric on $G$ we have
\begin{equation}\label{5.3}
\dd s^2_G=\delta_{ab}\theta^a_L\theta^b_L=\delta_{ab}\theta^a_R\theta^b_R\ ,
\end{equation}
where 
\begin{equation}\label{5.4}
\theta^a_R=D^a_b\theta^b_L\quad\mbox{for}\quad gI_ag^{-1}=:D_a^bI_b\ .
\end{equation}
From \eqref{5.4} one can see that left- and right-invariant objects are interchangable. 

The forms $\theta^a_L$ obey the Maurer-Cartan equations
\begin{equation}\label{5.5}
\dd_G\theta_L^a + \sfrac12\,f^a_{bc}\theta_L^b\wedge\theta_L^c=0
\end{equation}
and the same equations for $\theta^a_R$ with $f^a_{bc}\to -f^a_{bc}$. We introduce left- and right-invariant vector fields on $G$ dual to
$\theta^a_L$ and $\theta^a_R$,
\begin{equation}\label{5.6}
L_a\lrcorner\,\theta^b_L = \delta_{a}^b\quad\mbox{and}\quad R_a\lrcorner\,\theta^b_R = \delta_{a}^b\ ,
\end{equation}
which obey the equations
\begin{equation}\label{5.7}
[L_a, L_b] = f^c_{ab}L_c\quad\mbox{and}\quad [R_a, R_b] = - f^c_{ab}R_c
\end{equation}
and commute with each other.

\smallskip

\noindent
{\bf Adjoint representation.}  We consider the group $G=\sSU(N)$ in the {\it fundamental} (defining) representation. The center of $\sSU(N)$ is given by the matrices $\zeta\unit_N$, where $\zeta$ is the $N$-th root of unity, $\zeta^N=1$, i.e.
\begin{equation}\label{5.8}
Z(\sSU(N)) = \Z/N\Z =:\Z_N\ .
\end{equation}
Let us consider the left action of $G$ on itself,
\begin{equation}\label{5.9}
G\ni f\ \ \mapsto\ \ f^g:= gfg^{-1}\in G
\end{equation}
for $f,g\in G$. The maps \eqref{5.9} are {\it inner automorphisms} of $G$ denoted Inn$(G)$ and we have an isomorphism
\begin{equation}\label{5.10}
G/Z(G)\cong \mbox{Inn}(G)\ \Rightarrow\ \mbox{Inn}(\sSU(N))=\sSU(N)/\Z_N\ .
\end{equation}
The group $\sSU(N)/\Z_N$ is locally isomorphic to the group $\sSU(N)$ which is a $\Z_N$-cover of $\sSU(N)/\Z_N$. The well-known example is the group $\sSO(3)=\sSU(2)/\Z_2$.

The group $G^\prime :=\mbox{PU}(N)=\sSU(N)/\Z_N$ (projective unitary group) has no $N$-dimensional representations. The adjoint {\it action} \eqref{5.9} of $\mbox{PU}(N)$ on $\sSU(N)$ induces the action
\begin{equation}\label{5.11}
\Ad_g:\ \ T_e G\to T_e G, \quad \frg\ni\vph\ \mapsto\ g\vph g^{-1}\in \frg\ ,
\end{equation}
where $\frg = T_e G$ is the tangent space of $G$ at the origin $e$. Thus, the group $G^\prime =\mbox{PU}(N)$ has $(N^2-1)$-dimensional representation which is the {\it adjoint} representation of $\sSU(N)$. Matrices $D=(D^a_b)$ introduced in \eqref{5.4} are matrices of this representation $g\to D$. For  fields $\vph =\vph^aI_a$ in the adjoint representation we have
\begin{equation}\label{5.12}
\widetilde\vph = \widetilde\vph^aI_a:=g\vph g^{-1}=(gI_b g^{-1})\vph^b=D^a_b\vph^bI_a\ \ \Rightarrow\ \ \widetilde\vph^a=D^a_b\vph^b\ ,
\end{equation}
i.e. they are transformed with the matrices $D$. The metric \eqref{5.3} is invariant under these rotations.

\smallskip

\noindent
{\bf Rescaling.} Note that $\theta^a_L$ and $L_a$ are defined in terms of angle variables on $G$ and one can interpret the metric \eqref{5.3} as a metric with the length parameter $R_0$ fixed to unity. Consider now the dimensionless parameter $\sigma :=g^{-1}_*$ and rescale $\theta^a_L$ and $L_a$ as 
\begin{equation}\label{5.13}
\theta^a_L\ \to\ \widetilde\theta^a_L=\sigma\theta^a_L=g^{-1}_*\theta^a_L\quad\mbox{and}\quad
L_a\ \to\ \widetilde L_a=\sigma^{-1}L_a=g_*L_a\ ,
\end{equation}
where $g_*$ is the coupling parameter. We have
\begin{equation}\label{5.14}
\dd_G\widetilde\theta_L^a + \sfrac12\, g_*f^a_{bc}\widetilde\theta^b_L\wedge\widetilde\theta^c_L=0\ ,\quad [\widetilde L_a, \widetilde L_b] = g_*f^c_{ab}\widetilde L_c
\end{equation}
and the same formulae for $\widetilde\theta_R^a$, $\widetilde R_a$. This rescaling is equivalent to the rescaling $D=(D^a_b)\to \sigma D$ of matrices $D\in G^\prime$ representing matrices $g\in G$ under the homomorphism $G\to G^\prime$, with $G^\prime$ action on $\frg$.

For the rescaled metric we have
\begin{equation}\label{5.15}
\dd\widetilde s^2_G=\delta_{ab}\widetilde\theta^a_L\widetilde\theta^b_L=\sigma^2\delta_{ab}\theta^a_L\theta^b_L =:\widetilde g_{ab}\theta^a_L\theta^b_L\ ,
\end{equation}
\begin{equation}\label{5.16}
\Rightarrow\ \widetilde g_{ab}=\sigma^2\delta_{ab}=g_*^{-2}\delta_{ab}\ .
\end{equation}
From \eqref{5.16} it follows that the Lagrangian \eqref{2.5} for pure Yang-Mills fields with $v=0$ can be written as
\begin{equation}\label{5.17}
-\frac{1}{4g^2_*}\delta_{ab}\CF^a_{\mu\nu}\CF^{b\mu\nu}= -\frac{1}{4}\widetilde g_{ab}\CF^a_{\mu\nu}\CF^{b\mu\nu}\ ,
\end{equation}
i.e. $g_*$ defines a scale \eqref{5.13}-\eqref{5.16} on the algebra $\frg =\,$Lie$\,G=\,$Lie$\,G^\prime$. All fields in the adjoint representation of  SU($N$) are transformed in fact by the group PU$(N)=\sSU(N)/\Z_N$.

\smallskip

\noindent
{\bf Group $\hat G$.}  We consider the scaling factor $\sigma =g_*^{-1}$ and the group
\begin{equation}\label{5.18}
\hat G=\R^+\times G^\prime =\bigl\{\hat D=\sigma D,\ \sigma\in\R^+,\ D\in G^\prime\mid \hat D^T\hat D=\sigma^2\unit_{N^2-1}\bigr\}
\end{equation}
with $(N^2-1)\times (N^2-1)$ matrices $D$ for $G^\prime$  embedded into the orthogonal group SO$(N^2-1)$. Thus, we consider the conformal extension of the adjoint representation of $G=\sSU(N)$. In \eqref{3.4} we introduced such an extension for the fundamental representation of this group. 

The metric on $\hat G$ is given by formula
\begin{equation}\label{5.19}
\dd s^2_{\hat G}=\dd\sigma^2 + \sigma^2\dd s^2_{G^\prime}\ .
\end{equation}
Under a homomorphism of the group \eqref{3.4} into the group \eqref{5.18} we have a map of $\rho$ into $\sigma$. It is natural to consider $\sigma = \rho^N$ since SU$(N)$ is $N$-fold covering of $\sSU(N)/\Z_N$.

The adjoint representation of $G=\sU(1)$ is trivial and the above logic cannot be used. The Abelian case is essentially different from the non-Abelian one, since for it we have two multiplicative groups $\C^*$ with coordinates
\begin{equation}\nonumber
\zeta = \rho e^{\im\theta}\quad\mbox{and}\quad z=\sigma e^{\im\vartheta}
\end{equation}
The natural homomorphism $p: \C^*\to \C^*$ is
\begin{equation}\label{5.20}
\zeta \to z=\zeta^k\ ,\quad k\in\Z\setminus\{0\}\ .
\end{equation}
The condition that the radii of both groups U(1) decrease and increase synchronously leads to the condition $\sigma = \rho^k$ with $k=1,2,...\ $ but the case $k=N=1$ is preferable.

\smallskip

\noindent
{\bf Associated bundles.}  Consider the principal bundle $\widetilde P(M, \widetilde G) = M\times\widetilde G$ of conformal frames on the vector bundle $\widetilde E=E\otimes L_+$ introduced in Section 4. The fibre of $\widetilde P\to M$ over a point $x\in M$ is the group $\widetilde G_x$ of $N\times N$ matrices defined in \eqref{3.4}. These fibres are parametrized by the Stueckelberg field $\phi\in G$ and the scaling field $\rho\in\R^+$,
$\Phi =\rho\phi\in\widetilde G=\R^+\times G$. Fibres of $\widetilde E\to M$ are spaces $\widetilde E_x\cong \C^N$ of the fundamental representation of the group $\widetilde G_x$. The metric \eqref{3.2} on $\widetilde E$ is
\begin{equation}\label{5.21}
\widetilde h(\psi )=\rho^2 h(\psi )=\rho^2\psi^\+\psi
\end{equation}
and the metric on $\widetilde G$ is given in \eqref{3.5}.

Similarly, we consider the bundle of Lie algebras Ad$P=P\times_G\frg$, where $G$ acts on $\frg$ by adjoint action \eqref{5.11} and introduce on $\frg_x$ over $x\in M$ the metric
\begin{equation}\label{5.22}
\hat q(\varphi )=\sigma^2q(\varphi ) =\sigma^2(x) \delta_{ab}\vph^a\vph^b=\langle\hat\vph, \hat\vph\rangle
\end{equation}
for sections $\vph = \vph^aI_a$ and $\hat\vph = \sigma\vph$ of the bundle Ad$P$. We associate with this vector bundle the bundle $\hat P(M, \hat G)$ of conformal frames, where the group $\hat G$ is given in \eqref{5.18}. The scaling functions $\rho$ and $\sigma =g_*^{-1}$ are not independent, as we discussed above. The scaling function $\rho$ is a dynamical quantity governed by the Lagrangian \eqref{4.10}. From the proposed model it follows that $\sigma$ tends to zero if $\rho$ tends to zero, i.e. both the vector bundle $\widetilde E$ and the adjoint vector bundle Ad$\hat P$ disappear in the limit $\rho\to 0$ ($g_*\to\infty$) since their fibres shrink to a point. Considering conformally rescaled metrics $\widetilde g=\Omega^2g$ on $M$, $\widetilde h =\rho^2h$ on $\widetilde E$ and $\hat q=\sigma^2q$ on Ad$\hat P$, we enter the region of real-space renormalization group already at the classical level due to assignment of geometric status to the Higgs fields.

\section{Building Models}

\noindent
{\bf Proposed ideas.} Before proceeding to the discussion of the Standard Model (SM) with the structure group $\sSU(3)\times\sSU(2)\times\sU(1)$, we summarize the ideas discussed.

\begin{itemize}
\item All fields entering in Lagrangian \eqref{4.10} are related to the geometry of fibre bundles $\widetilde P(M,\widetilde G)$, $\widetilde E$, $\hat P(M, \hat G)$ and  Ad$\hat P$.

\item The $G$-valued Stueckelberg field $\phi$ parametrizes frames on the complex vector bundle $E$ and these frames are dynamical if Lagrangian contains the term $\rho^2\tr (\nabla\phi)^2$. Scaling function $\rho$ sets the value of effective mass of gauge bosons.

\item We have Yang-Mills-Higgs theory invariant under automorphisms Aut$_GE$ (gauge symmetry) and the Higgs field $\Phi =\rho\phi\in\widetilde G$. If we do not add the term $\rho^2\tr (\nabla\phi )^2$ to the Lagrangian, then gauge bosons are massless.

\item The scaling function $\rho$ as well as the scale connection $a=a_\mu\dd x^\mu$ enter the Lagrangian both in the massive and in the massless case.

\item Scaling function $\rho (x)$ defines ``size'' of fibres $\widetilde G_x$ of the bundle $\widetilde P(M, \widetilde G)$ of frames on the vector bundle $\widetilde E$ and the inverse coupling $\sigma (x)$ defines ``size'' of fibres $\hat G_x$ of the bundle $\hat P(M, \hat G)$ of frames on the adjoint bundle Ad$\hat P$ of algebras.

\item If $\rho , \sigma$ tend to zero, then fibres of all gauge bundles shrink to a point, and there are no gauge fields in the region where $\rho$ and $\sigma$ are equal to zero.
\end{itemize}

These ideas form a hard core of the proposed research program.

\smallskip

\noindent
{\bf Geometry and matter.}  We consider two types of objects: those that define geometry and those that define matter.

{\it Geometry:}
\begin{itemize}
\item Orthonormal (co-)frame  $\Theta^\mu$ and Weyl conformal factor $\Omega$ on the cotangent bundle $T^*M$ over curved 4-manifold $M$ define conformal geometry of $M$.

\item A $G$-frame $\phi$ and conformal factor $\rho$ on Hermitian vector bundle $\widetilde E\to M$ define geometry of $\widetilde E$.

\item Connections $a, \CA$ and the inverse running coupling $\sigma=g_*^{-1}$ define a frame on the adjoint bundle. In particular, a metric on $\hat P(M,\hat G)$ can be written in the form
\begin{equation}\label{6.1}
\dd s^2_{\hat P}=\Omega^2\eta_{\mu\nu}\Theta^\mu\Theta^\nu + \Theta^\sigma\Theta^\sigma + \sigma^2\delta_{ab}\Theta^a\Theta^b\ ,
\end{equation}
where $\Theta^\mu$'s define a metric on $M$ and
\begin{equation}\label{6.2}
\Theta^\sigma =\dd\sigma + N\sigma a_\mu\dd x^\mu\ ,\quad \Theta^a = \theta^a_R - D^a_b\CA^b_\mu \dd x^\mu
\end{equation}
define a metric on fibres $\hat G_x$ for $\theta^a_R$ and $D^a_b$ introduced in Section 5. Similar metric with $\rho$ instead $\sigma$ can be written on $\widetilde P(M,\widetilde G)$. 
\end{itemize}

Thus, geometry is defined by {\it frames} and the associated {\it metrics}.

{\it Matter:}
\begin{itemize}
\item Matter fields are given by {\it sections} of tensor products of vector bundles over $M$.
\end{itemize}

In the Standard Model we have a complex vector bundle $\CE =E_{\C^3}\otimes E_{\C^2}\otimes E_{\C}$ associated with the group $\sSU(3)\times\sSU(2)\times\sU(1)$. Quarks and leptons are {\it sections} of the bundle $\CE$ tensored with the spinor bundle over $M$. They are matter fields.

Sections of vector bundles do not affect the geometry, so they can have any value, including zero. For example, a vector field $W$ on $M$ can be written in local coordinates as $W=W^\mu\partial_\mu\in TM$, where $W^\mu$ are any functions of $x\in M$. Zero values of $W^\mu$ simply mean that there is no vector field, and the same can be said for the fields of quarks and leptons. At the same time, if scaling functions $\Omega$ or $\rho$ are equal to zero at some points of $M$ or in some region of $M$, then this radically changes the geometry of the manifold $M$ and bundles over it.

\smallskip

\noindent
{\bf Fermions.} Let $\psi$ be a fermionic field with value in the complex vector bundle $E\to M$. It is a section of the bundle $E$ tensored with the spinor bundle over $M=\R^{3,1}$. The standard Lagrangian density for $\psi$ has the form
\begin{equation}\label{6.3}
\CL_\rmF=\bar\psi\im\gamma^\mu\nabla_\mu\psi - m\bar\psi\psi\ , 
\end{equation}
where $\gamma$-matrices satisfy the anticommutation relations $\{\gamma^\mu , \gamma^\nu\} = - 2\eta^{\mu\nu}\unit_4$, $\bar\psi =\psi^\+\gamma^0$ and $m$ is the mass of the fermion $\psi$. Note that we can add to $\psi$ a flavour index and sum over it in \eqref{6.3}.

In \eqref{3.1}-\eqref{3.4} we introduced dressed fields
\begin{equation}\label{6.4}
\tilde\psi =\Phi\psi\quad\mbox{for}\quad\Phi=\rho\phi\in\tilde G=\R^+\times G
\end{equation}
taking value in the bundle $\tilde E$ and in  \eqref{4.5}-\eqref{4.8} we coupled $\tilde\psi$ with a scale connection $a=a_\mu\dd x^\mu$. For such fields we have 
\begin{equation}\label{6.5}
\widetilde\CL_\rmF=\bar{\tilde\psi}\im\gamma^\mu\bigl(\nabla_\mu + a_\mu\bigr)\tilde\psi - m\bar{\tilde\psi}\tilde\psi=\rho^2\CL_\rmF + \rho^2
\bigl(a_\mu + \partial_\mu\ln\rho\bigr) \bar\psi\im\gamma^\mu\psi   \ , 
\end{equation}
where $\CL_\rmF$ is given in \eqref{6.3} and $\nabla_\mu =\partial_\mu + \CA_\mu$. Note that $\widetilde \CL_\rmF=0$ for $\rho =0$ as expected.

\smallskip

\noindent
{\bf Higgs fields.} What about the Higgs field - is it matter or geometry? It is well known that the prototype for the Higgs model was the Ginzburg-Landau model of superconductivity. The Lagrangian of this Abelian model with $G=\sU(1)$ has the form \eqref{4.10} with $\phi\in\sU(1)$, $\rho\in [0,\infty )$, $a=0=f$ and 
\begin{equation}\label{6.6}
\rho^2_0=\gamma^2(T_c - T)\ ,
\end{equation}
where $T$ is temperature and $T_c$ is the transition temperature. The order parameter field
\begin{equation}\label{6.7}
\psi=\rho\phi\in\C
\end{equation}
is the ``condensate'' of Cooper pairs of electrons, i.e. $\psi$ is a composite {\it matter} field with $\rho^2=\bar\psi\psi$ indicating the fraction of electrons that have condensed into a superfluid. The Higgs field $\psi$ in this case is a {\it section} of the complex line bundle $E_\C$ associated with the electromagnetic $\sU(1)$-bundle. From \eqref{6.6} it can be seen that at $T>T_c$ the minimum of the potential is possible only at $\rho =0$, i.e. superconductivity disappear. In the Ginzburg-Landau model the field \eqref{6.7} is not related to geometry, electric charge does not depend on $\rho$ and equality $\psi =0$ simply means the absence  of Cooper pairs. 

Generalizing the Abelian case \eqref{6.7}, the Higgs field in SM was introduced as a section $\psi$ of the complex vector bundle $E_{\C^2}$, i.e. as matter. It is fundamental and not composite like Cooper pairs. In color superconductivity, there are composite matter Higgs fields (condensate of pairs of quarks). However, the fundamental scalar field should be considered as a geometric one. We showed that it is a conformal frame on the Hermitian vector bundle $E_{\C^N}$ associated with the group $G$ which is $\sSU(N)$ for $N>1$ and $\sU(1)$ for $N=1$. They are parametrized by fields $\Phi_N=\rho_N\phi_N\in\R^+\times G$ and give mass to gauge bosons only if the Stueckelberg fields $\phi_N$ are included in the Lagrangian. Furthermore, as $\rho_N$ tends to zero, the fibres of all gauge bundles shrink to a point, and gauge fields disappear. In this limit there is nothing to ``gauge''. 

Next, we will discuss fields $\Phi_N$ in the Standard Model. But before proceeding to this discussion, let us pay attention to two important features of the field \eqref{6.7} in the theory of superconductivity. Firstly, the field $\psi$ exists only in a compact region $S$ of $\R^3$ filled by a superconductor. Secondly, the minimum of potential energy is reached for $\rho_0^2$ from \eqref{6.6} depending on external parameters. I propose to consider Higgs-type fields $\Phi_N$ in particle physics also having such properties and also propose to consider $\rho_0$ as a function depending on coordinates $x\in M$. To begin with, we specify examples of functions supported on some subspaces $S$ of $\R^3$, compact or noncompact.\footnote{The program for the study of gauge theories with fields given on regions $S\subset\R^3$ was proposed in \cite{Popov}. Here, we continue to work on the implementation of this program.}

\smallskip

\noindent
{\bf Functions supported on $S$.} Suppose that $f: X\to\R$ is a real-valued function whose domain is an arbitrary set $X$. The {\it set-theoretical support} of $f$, written as supp$(f)$, is the set of points in $X$, where $f$ is non-zero:
\begin{equation}\label{6.8}
\mbox{supp}(f) = \bigl\{x\in X\mid f(x)\ne 0\bigr\}\ .
\end{equation}
For example, a {\it characteristic function} of a subset $S\subset X$ is the function
\begin{equation}\label{6.9}
\unit_S:\ \ X\to \{0,1\}\ ,\  \quad \unit_S(x)=\left\{\begin{array}{l}1\ \mbox{for}\ x\in S\\0\ \mbox{for}\ x\not\in S\end{array}\right .\ .
\end{equation}
The function $\unit_S$ indicates whether $x\in X$ belongs to $S$ or not. Obviously, $1{-}\unit_S$ is the characteristic function of the complement $X{\setminus}S$. When $X$ is a topological space, the support of $f$ is defined as a closure in $X$ of the subset in $X$ where $f$ is non-zero,
\begin{equation}\label{6.10}
S=\mbox{supp}(f) = \mbox{cl}_X\bigl(\{x\in X\mid f(x)\ne 0\}\bigr)\ .
\end{equation}
Below we will consider functions on $X=\R^3$ with coordinates $x\in\R^3$ and use $r^2=(x^1)^2+(x^2)^2+(x^3)^2$.

Let $S$ be a compact (closed and bounded) embedded submanifold of a Euclidean space. Real-valued compactly supported functions on a Euclidean space are called {\it bump functions}. We will consider a 3-dimensional submanifold $S$ in $\R^3$ as the closed 3-ball $\bar B^3_R(0)$ of radius $R$ centered at $x=0$,
\begin{equation}\label{6.11}
S=\bar B^3_R(0) = \bigl\{x\in \R^3\mid r^2\le R^2\bigr\}\ .
\end{equation}
As an example of a bump function we consider a function
\begin{equation}\label{6.12}
\chi^{}_{<R}=\left\{
\begin{array}{r}\exp\left (\frac{r^2}{r^2-R^2}\right )\ \mbox{for}\ r^2<R^2\\
0\qquad\qquad\mbox{for}\ r^2\ge R^2\end{array}\right .
\end{equation}
which can be written as
\begin{equation}\label{6.13}
\chi^{}_{<R}=\exp\left (\frac{r^2}{r^2-R^2}\right )\, \unit^{}_{\{r<R\}}\ ,
\end{equation}
where $\unit^{}_{\{r<R\}}$ is the characteristic function of the open ball $B^3_R(0)$. We have $\chi^{}_{<R}\to 0$ for $r{\nearrow}R$ (the limit from below). 

The space of bump functions is closed under the sum, product or convolution of two bump functions. Any differential operator with smooth coefficients, when applied to a bump function, will produce another bump function. For example, $\chi^{N}_{<R}$ for $N\in\Nbb$ and $f\chi^{}_{<R}$ for a smooth function $f$ are again bump functions.

As an example function having noncompact support we consider a function
\begin{equation}\label{6.14}
\zeta^{}_{>L}=\exp\left (-\frac{L^2}{r^2-L^2}\right )\, \unit^{}_{\{r>L\}}
\end{equation}
such that $\zeta^{}_{>L}\to 1$ for $r\to\infty$ and $\zeta^{}_{>L}\to 0$ for $r{\searrow}L$ (the limit from above). Note that we can multiply $\zeta^{}_{>L}$ by a smooth function and again have a function supported on $S=\bigl\{x\in \R^3\mid r\ge L\bigr\}$. As a last example, we introduce a bump function
\begin{equation}\label{6.15}
\zeta^{}_{>L}\chi^{}_{<R}=\exp\left (\frac{r^2}{r^2-R^2}-\frac{L^2}{r^2-L^2}\right )\, \unit^{}_{\{L<r<R\}}\ ,
\end{equation}
where $\unit^{}_{\{L<r<R\}}=\unit^{}_{\{L<r\}}\unit^{}_{\{r>R\}}$ for $L<R$. Different products of positive interger powers of these functions also give a bump function on $S=\bigl\{x\in \R^3\mid L\le r\le R\bigr\}$, e.g. $\zeta^{N_1}_{>L}\chi^{N_2}_{<R}$.

\smallskip

\noindent
{\bf Running couplings.} We turn to discussing the construction of models using the geometric fields $\rho_N$ and $\phi_N$ with $N=1,2,3,...$. Any constructions must be based on the facts established at the moment in experiments and theoretical studies.

1) QED, $N=1, G=\sU(1)$: It is known that photons are massless and hence there should be no term $\rho_1(\nabla\phi_1)^2$ in the Lagrangian. There is the energy scale at which the coupling parameter $g^{\sf eff}_1$ becomes infinite (Landau pole). We denote by $L$ the length scale such that $g^{\sf eff}_1\to\infty$ for $r{\searrow}L$ and this length $L$ can be smaller than the Plank length. This behaviour of $g^{\sf eff}_1$ is usually explained by vacuum polarization by virtual electron-positron pairs. 

If we assume the reality of this Landau pole and interpret $g^{\sf eff}_1$ as a function inverse to the $k$-th power $\rho_1^k\  (k{\ge}1)$ of the scalar field $\rho_1$ defining radii of U(1)-fibres in the bundle $\widetilde P(\R^{3,1}, \widetilde U(1))$, then this simply means that electromagnetic fields are defined only outside the ball $B^3_L(0)=\{x\in\R^3\mid r<L\}$ since $\rho_1\to 0$ for $r{\searrow}L$. An example of functions with such behaviour is the function $\zeta_{>L}$ from \eqref{6.14}.

2) QCD, $N=3, G=\sSU(3)$: Gluons are considered massless, but there are a number of indications that they can be massive (see e.g. \cite{Pel} and references therein). It is known that the coupling parameter $g^{\sf eff}_3$ tends to zero at small $r$ (asymptotic freedom) and we assume that $g^{\sf eff}_3\to 0$ for $r{\searrow}L$. It is also known that $g^{\sf eff}_3$ tends to infinity at some scale $R_3$, which specifies the size of hadrons, and this behaviour leads to confinement. 

From a mathematical point of view, confinement in QCD means that all functions defining quarks and gluons have a {\it compact support} on the bag  \eqref{6.11}. If we regard this behaviour as a consequence of the geometry of $\sSU(3)$-bundles and interpret $g^{\sf eff}_3$ as the inverse of the scale function $\sigma_3$, then the QCD gauge bundles are defined only in the region $S=\{x\in\R^3\mid L\le r\le R_3\}$. An example of functions $g^{\sf eff}_3$ with such behavior is given by $g^{\sf eff}_3=g_3\zeta_{>L}^3\chi^{-3}_{<R_3}$ 
for $\zeta_{>L}$ and $\chi_{<R}$ from \eqref{6.12}-\eqref{6.14}. Such a behavior of $g^{\sf eff}_3$ is usually explained by a cloud of virtual gluons and quark-antiquark pairs. These virtual particles live inside the ball $\bar B^3_{R_3}(0)$ with $R_3\gg L$ and tend to accumulate near its boundary from the inside. Note that the Higgs potential in QCD is usually chosen to be zero.

3) Weak $\sSU(2), N=2$: It is known that SU(2) coupling parameter $g^{\sf eff}_2$ tends to infinity in the infrared region, at $r{\nearrow}R_2$ for some  $R_2>R_3$, i.e. SU(2) theory is confining. Also, there is the Higgs potential with the expectation value $v=:\rho_H$ for $\rho_2$  corresponding to some $R_H$ and this expectation value cuts off the growth of $g^{\sf eff}_2$ since
\begin{equation}\label{6.16}
L\ll R_H\ll R_3\ll R_2\ .
\end{equation}
The Higgs mechanism in the electroweak theory is turned off if $R_H>R_2$. Then both gauge groups SU(3) and SU(2) are asymptotically free and confining (see e.g. \cite{Berger, Tong} and references therein). In fact, the electroweak SU(2)$\times$U(1) case is a mixture of Abelian and non-Abelian behaviour of coupling parameters. Phenomenologically this is described in the Standard Model. However, one would like to get a clearer mathematical understanding and this requires additional efforts.

\smallskip

\noindent
{\bf Order parameter fields.} The introduced scaling parameters $\rho_N\ (N=1,2,3,... )$ are the order parameters specifying the points of ``phase transitions'' between different states. This $\rho\ ({=}\rho_N)$ defines sizes of fibres in the bundle $\widetilde P(M, \widetilde G)$ and $\widetilde E$, and for $\rho\to 0$ we obtain an empty space $M=\R^{3,1}$ without gauge and fermionic fields. Recall that the metric on the group $\widetilde G$ is 
\begin{equation}\label{6.17}
\dd s^2_{\widetilde G} = \dd\rho^2 + \rho^2\dd s^2_{G}
\end{equation}
and $\rho$ is the ``radius'' of the group $G_\rho$ in $\widetilde G\cong\R^+\times G$. For $\rho\to 0$ the group $G_\rho$ shrinks to a point and for $\rho\to\infty$ we have $g_\ast\to 0$ (asymptotic freedom) since the Lie algebra of $G_\infty$ is Abelian, e.g. SU(2)$_\rho = S^3_\rho\to\R^3$ for $\rho\to\infty$. Thus $\rho =0$ and $\rho =\infty$ are the critical points of the theories under consideration.

\smallskip

\noindent
{\bf Bag models.} It is believed that quarks and gluons can never be liberated from hadrons since the force between color charges increases with distance. The QCD Lagrangian is considered as a sum of Yang-Mills part
\begin{equation}\label{6.18}
\CL^{}_{\sf YM}=\frac{1}{4g^2_3}\,\tr (\CF_{\mu\nu}\CF^{\mu\nu})\ ,
\end{equation}
with $G=\sSU(3)$ and the coupling $g_3>0$ and the quark part \eqref{6.3} where a flavour index is hidden.

All hadrons are composed of quarks, and in 1974 the bag model was proposed to describe hadrons \cite{MIT}. This is a phenomenological model in which confinement of quarks and gluons is postulated by imposing boundary conditions such that all fields vanish outside the bag $S=\bar B^3_R(0)$. In this model, the Lagrangian density is chosen in the form
\begin{equation}\label{6.19}
\CL^{}_{\sf bag}=(\CL^{}_{\sf YM} + \CL^{}_{\rmF} -\Lambda )\unit_S + \mbox{boundary~terms~on~} \partial S\ ,
\end{equation}
where the constant $\Lambda >0$ is the vacuum energy density inside of $S$ (a local cosmological constant), $\unit_S$ is the characteristic function \eqref{6.9} and hence $\CL^{}_{\sf bag}(t, x)=0$ for $x\not\in S$. Note that the effective coupling in \eqref{6.19} is
\begin{equation}\label{6.20}
\frac{1}{g^2_{\sf eff}}=\frac{1}{g^2_3}\, \unit_S =\left\{\begin{array}{l}\frac{1}{g^2_3}\,\ \mbox{for}\ x\in S \\[3mm]
0\ \mbox{for}\ x\not\in S\end{array}\right . ,
\end{equation}
i.e. $g_{\sf eff}=\infty$  outside the bag $S$. It is this behavior that is responsible for confinement.

As a next step, a soliton bag model was proposed \cite{FrLee}, where $g_3^{-2}\unit_S$ in \eqref{6.19} was replaced by a function $g^{-2}_{\sf eff}=\sigma^2(\rho )$ depending on the dynamical scalar field $\rho$ included in the extended Lagrangian with terms
\begin{equation}\label{6.21}
\partial_\mu\rho\,\partial^\mu\rho + V(\rho )
\end{equation}
and coupling of $\rho$ with fermionic fields $\psi$. Then the field equations for $\rho$ contain quark fields and it has been shown that there are solutions such that $\rho , \sigma\to 0$ for $r{\nearrow}R$, where $R$ is the radius of the bag $S=\bar B^3_R(0)$ \cite{FrLee}. In fact, this model smoothes all functions of type $f\unit_S$ into bump functions $f_S$ supported on $S$ and shows that such solutions $\CA_S, \psi_S$ exist. This explains confinement, but this model was also phenomenological since  the origin of the fields $\rho$ and $\sigma$ was unknown. The search of more fundamental models of confinement continued, other models were proposed. A description of the most popular models can be found e.g. in \cite{Green}. All of these models, over more than 40 years of efforts, have failed to produce satisfactory results for QCD. Therefore, it makes sense to return to bag models using the ideas suggested above in this paper. 

\smallskip

\noindent
{\bf Dressed quarks.} Both the MIT bag model \cite{MIT} and its soliton bag generalization \cite{FrLee} contain the quark Lagrangian  \eqref{6.3} with fields $\CA_0$ and $\psi_0$. The quarks, which determine the quantum numbers of hadrons, have small masses and are called valence or {\it current}  quarks $\psi_0$. Apart from these, hadrons contain an infinite number of virtual gluons and quark-antiquark pairs. Current quarks surrounded by a cloud of these virtual particles form {\it constituent} quarks $\psi$ with measurable masses $m$ which are much larger than those of current quarks $\psi_0$. We can identify $\psi$ with the {\it dressed} current quarks $\psi_0$ via the maps \cite{McM}
\begin{equation}\label{6.22}
\psi_0\mapsto\psi = \phi\psi_0\quad\mbox{and}\quad\CA_0\mapsto\CA = \phi\CA_0\phi^\+ + \phi\dd\phi^\+
\end{equation}
discussed in Section 2. Here $\phi\in G$ is the Stueckelberg field and the Lagrangian  \eqref{6.3} contains the dressed fields  \eqref{6.22}.

Note that virtual particles generate not only $\phi\in G$ but also the scalar field $\rho\in\R^+$ discussed in Section 3. Therefore, we should pass from the field $\psi$ to the field  $\tilde\psi =\rho\psi$ with the addition of the scale connection $a=a_\mu\dd x^\mu$. Then for fermions we obtain the Lagrangian \eqref{6.5}, where $\CA$ has the form \eqref{6.22}. The field $\Phi = \rho\phi\in\R^+\times G$ is generated by the ``sea'' of virtual particles around quarks inserted into the bag $S=\bar B^3_R(0)$. 

\smallskip

\noindent
{\bf Confinement.} Let us consider the non-Abelian case with $G=\sSU (N)$ for $N>1$. In the potential energy $V(\rho )$ from the Lagrangian 
\eqref{4.10}, we leave only the $\rho$-independent term
\begin{equation}\label{6.23}
-\Lambda\chi_S\ ,
\end{equation}
which is a smoothed version of the term $-\Lambda\unit_S$ from \eqref{6.19}. Here $\chi_S$ is a bump function on the 3-ball $S=\bar B^3_R(0)$ and $\Lambda\ge 0$ is the energy density required to create a bubble $S$ with quarks. In \cite{Popov} we interpreted this term as the Dirichlet energy density associated with an embedding of a 3-ball in $\R^3$. In accordance with the discussion of Section 5, we set $g_*(\rho )=g_N\rho^{-N}$, where $g_N$ is a constant. Then, as a bosonic part of the Lagrangian density, we have
\begin{equation}\label{6.24}
\widetilde\CL^{}_{\rm B}=\frac{\rho^{2N}}{4g^2_N}\tr (\CF_{\mu\nu}\CF^{\mu\nu}){-}\frac{1}{4}f_{\mu\nu}f^{\mu\nu}{-}\frac{\varkappa\rho^2}{2}\tr (\nabla_\mu\phi)^\+\nabla^\mu\phi{-}\frac{N}{2}(\partial_\mu\rho{+}a_\mu\rho )(\partial^\mu\rho{+}a^\mu\rho ){-}\Lambda\chi_S.
\end{equation}
The Lagrangian density for fermions was introduced in \eqref{6.5},
\begin{equation}\label{6.25}
\widetilde\CL^{}_{\rmF}=\rho^2\CL_\rmF+\rho (\partial_\mu\rho + a_\mu\rho )\bar\psi\im\gamma^\mu\psi \ ,
\end{equation}
where $\CL_\rmF$ is given in  \eqref{6.3}, and for the full Lagrangian $\widetilde\CL$ we have
\begin{equation}\label{6.26}
\widetilde\CL =\widetilde\CL^{}_{\rm B}+\widetilde\CL^{}_{\rmF}\ .
\end{equation}
In \eqref{6.24} we  choose $\varkappa =0$ for massless and $\varkappa =1$ for massive gauge bosons.

The Lagrangian \eqref{6.26} is a modification of Lagrangians from \cite{MIT} and \cite{FrLee} and it is natural to expect that it will give solutions with fields supported on $S$. We will not attempt to show this as it is not the purpose of this paper. Consider only one possible chain of reasoning. Non-Abelian gauge bosons cannot propagate freely in space far from sources, unlike photons. Therefore, in empty space without quarks $\psi$, we have $\rho =0=\Lambda$ and all terms of the Lagrangians \eqref{6.24}-\eqref{6.26} are equal to zero. We assume that $\rho$ vanishes outside the bag $S$ and introduce the functions
\begin{equation}\label{6.27}
\rho_S:=\chi^{}_{<R}\zeta^{-1}_{>L}\quad\mbox{with}\quad R\equiv R_N\quad\mbox{and}\quad\chi :=\rho\rho_S^{-1}\ ,
\end{equation}
where $\chi^{}_{<R}$ and $\zeta^{}_{>L}$ are given in \eqref{6.12}-\eqref{6.14}, and choose
\begin{equation}\label{6.28}
a_\mu = -\partial_\mu \ln \rho_S\quad\Rightarrow\quad f_{\mu\nu}=0\ .
\end{equation}
Then the Lagrangian \eqref{6.26} will read as
\begin{equation}\label{6.29}
\widetilde\CL{=}\frac{\chi^{2N}\rho^{2N}_S}{4g_N^2}\tr (\CF_{\mu\nu}\CF^{\mu\nu}){-}\frac{\varkappa\chi^{2}\rho^{2}_S}{2}
\tr (\nabla_\mu\phi)^\+\nabla^\mu\phi{-}\frac{N\rho^2_S}{2}\partial_\mu\chi\partial^\mu\chi
{-}\Lambda\chi_S{+}\chi^{2}\rho^{2}_S\CL_\rmF{+}\chi(\partial_\mu\chi )\rho^{2}_S\bar\psi\im\gamma^\mu\psi .
\end{equation}
Varying this Lagrangian with respect to $\chi$, we obtain the equation
\begin{equation}\label{6.30}
\partial_\mu \bigr(\rho^{2}_S\partial^\mu\chi\bigl) + \chi\left\{\frac{\rho^{2N}_S}{2g_N^2}\tr (\CF_{\mu\nu}\CF^{\mu\nu}){-}\frac{1}{N}\Bigl[\partial_\mu(\rho^{2}_S\bar\psi\im\gamma^\mu\psi ){-}2\rho^{2}_S\CL_\rmF{+}\varkappa\rho^{2}_S\,\tr(\nabla_\mu\phi)^\+\nabla^\mu\phi \Bigr]\right\}=0.
\end{equation}
The obvious solution is $\chi=0$. For $\chi =1$, we obtain a relation between gluon and quark fields, which can always be fulfilled.

\smallskip

\noindent
{\bf $W$-bosons.} The case of $G=\sSU(2)$ was discussed around \eqref{6.16}. The $\sSU(2)$ case is also confining and asymptotically free with $\rho_S$ given by \eqref{6.27} with $R=R_2$ and $g_2^{\sf eff}\sim g_2\rho_S^{-2}$. However, the SM mass scale of the weak interaction is determined not by $R_2$ but by the vacuum expectation value $v=\rho^{}_H$ of the scalar $\rho$ corresponding to the radius $R_H\ll R_2$. This expectation value $\rho^{}_H$ also cuts off the infrared growth of the $\sSU(2)$ coupling constant $g_2^{\sf eff}$ if we add to 
\eqref{6.24} with $N=2$ the potential
\begin{equation}\label{6.31}
-\frac{\lambda^{}_S}{4}\,(\rho^2 - \rho_H^2)^2\ ,
\end{equation}
where $\lambda^{}_S$ is a bump function on $S=\bar B^3_{R_2}(0)$. For $r\sim R^{}_H$ this $\lambda^{}_S$ is approximately constant as well as $\rho^{}_S$ and  $g^{\sf eff}_2$ which bring us back to the standard Higgs mechanism. Note, however, that the Higgs potential of the form \eqref{6.31} is quite arbitrary and inexplicable from a mathematical point of view. There is no justification for it other than the conformity with experimental data.

\smallskip

\noindent
{\bf Abelian case.} For $N=1$ and $G=U(1)$ the Lagrangian density is 
\begin{equation}\label{6.32}
\widetilde\CL = {-}\frac{\rho^{2k}}{4g_1^2}\CF_{\mu\nu}\CF^{\mu\nu}{-}\frac{1}{4}f_{\mu\nu}f^{\mu\nu}{-} 
 \rho^2(a_\mu{+}\partial_\mu\ln\rho)(a^\mu{+}\partial^\mu\ln\rho){+}\rho^{2}\CL_\rmF{+}\rho^2(a_\mu{+}\partial_\mu\ln\rho)\bar\psi\im\gamma^\mu\psi ,
\end{equation}
where $\CF =\dd\CA$ for real-valued $\CA$. Let us assume that the function $\rho$ behaves as $\zeta_{>L}$ from \eqref{6.14} and $k=1=N$. Choosing  $\rho=\chi\zeta_{>L}$ and $a=-\dd\ln\zeta_{>L}$ we obtain for $\chi$ the equation
\begin{equation}\label{6.33}
\partial_\mu \bigr(\zeta_{>L}^{2}\partial^\mu\chi \bigl)- \chi\zeta_{>L}^{2}\left\{\frac{1}{2g_1^2}\,\CF_{\mu\nu}\CF^{\mu\nu}
+ 2(\partial_\mu\ln\zeta_{>L}^{})\bar\psi\im\gamma^\mu\psi +  \partial_\mu (\bar\psi\im\gamma^\mu\psi )
+2\CL_\rmF\right\}=0\ .
\end{equation}
Then for $\chi =1$ we obtain a relation between bosonic and fermionic fields. Note that $\zeta_{>L}^{}\simeq 1$ for $r\ge R_3$ (size of hadrons) since $L\ll R_3$. 

In conclusion, we emphasize that the models building is not the main objective of this paper. Our intention was to give a geometric meaning to the fields $\rho_N, \phi_N$, $g^{\sf eff}_N$ and to formulate a program for further research. Studying specific models requires additional efforts.

\bigskip 

\noindent {\bf Acknowledgments}

\noindent
I am grateful to Tatiana Ivanova for stimulating discussions.
This work was supported by the Deutsche Forschungsgemeinschaft grant LE~838/19.

%\newpage


\begin{thebibliography}{99}
\bibitem{MIT}
A.~Chodos, R.L.~Jaffe, K.~Johnson, C.B.~Thorn and V.F.~Weisskopf, \\
``New extended model of hadrons,''
Phys. Rev. D  \textbf{9} (1974)  3471.

\bibitem{FrLee}
R.~Friedberg and T.D.~Lee,
``QCD and the soliton model of hadrons,''\\
Phys. Rev. D \textbf{18} (1978) 2623.

\bibitem{Popov}
A.D.~Popov,
``Yang-Mills-Stueckelberg theories, framing and local breaking of symmetries,''
[arXiv:2110.00405 [hep-th]].

\bibitem{McM}
M.~Lavelle and D.~McMullan,
``Constituent quarks from QCD,'' \\
Phys. Rept. \textbf{279} (1997) 1
[arXiv:hep-ph/9509344 [hep-ph]].

\bibitem{Franc}
J.~Fran\c{c}ois,
``Bundle geometry of the connection space, covariant Hamiltonian formalism, the problem of boundaries in gauge theories, and the dressing field method,''\\
JHEP \textbf{03} (2021) 225
[arXiv:2010.01597 [math-ph]].

\bibitem{Berg}
P.~Berghofer, J.~Fran\c{c}ois, S.~Friederich, H.~Gomes, G.~Hetzroni, A.~Maas and R.~Sondenheimer,
{\it Gauge symmetries, symmetry breaking, and gauge-invariant approaches},\\ Cambridge Elements, Cambridge University
Press, 2021.

\bibitem{Stueck}
E.C.G.~Stueckelberg, ``Die Wechselwirkungs Kraefte in der Elektrodynamik und in der Feldtheorie der Kernkraefte (I),''
%The interaction forces in electrodynamics and in the field theory of nuclear forces (I)
Helv. Phys. Acta \textbf{11} (1938) 225;\\
E.C.G.~Stueckelberg, ``Die Wechselwirkungs Kraefte in der Elektrodynamik und in der Feldtheorie der Kernkraefte (II),''
%The interaction forces in electrodynamics and in the field theory of nuclear forces (II)
Helv. Phys. Acta \textbf{11} (1938) 299.

\bibitem{BEH}
F.~Englert and R.~Brout,
``Broken symmetry and the mass of gauge vector mesons,''\\
Phys. Rev. Lett. \textbf{13} (1964) 321;\\
P.W.~Higgs,
``Broken symmetries and the masses of gauge bosons,''\\
Phys. Rev. Lett. \textbf{13} (1964) 508.

\bibitem{Ruegg}
H.~Ruegg and M.~Ruiz-Altaba,
``The Stueckelberg field,''\\
Int. J. Mod. Phys. A \textbf{19} (2004) 3265
[arXiv:hep-th/0304245 [hep-th]].

\bibitem{GSW}
J.L.~Gervais, B.~Sakita and S.~Wadia,
``The surface term in gauge theories,''\\
Phys. Lett. B \textbf{63} (1976) 55.

\bibitem{DF}
W.~Donnelly and L.~Freidel,
``Local subsystems in gauge theory and gravity,''\\
JHEP \textbf{09} (2016) 102
[arXiv:1601.04744 [hep-th]].

\bibitem{Str}
A.~Strominger,
{\it Lectures on the infrared structure of gravity and gauge theory,}\\
Princeton University Press, Princeton, 2018.

\bibitem{Kob} S. Kobayashi, {\it Transformation groups in differential geometry}, Springer, Berlin, 1972.

\bibitem{Scholz}
E.~Scholz,
``The unexpected resurgence of Weyl geometry in late 20-th century physics,''
Einstein Stud. \textbf{14} (2018) 261
%doi:10.1007/978-1-4939-7708-6\_11
[arXiv:1703.03187 [math.HO]].

\bibitem{Pel}
M.~Pel\'aez, U.~Reinosa, J.~Serreau, M.~Tissier and N.~Wschebor,\\
``A window on infrared QCD with small expansion parameters,''\\
Rept. Prog. Phys. \textbf{84} (2021) 124202
[arXiv:2106.04526 [hep-th]].

\bibitem{Berger}
J.~Berger, A.J.~Long and J.~Turner,
``Phase of confined electroweak force in the early Universe,''
Phys. Rev. D \textbf{100} (2019)  055005
[arXiv:1906.05157 [hep-ph]].

\bibitem{Tong}
N.~Lohitsiri and D.~Tong,
``If the weak were strong and the strong were weak,''\\ 
SciPost Phys. \textbf{7} (2019) 059
[arXiv:1907.08221 [hep-th]].

\bibitem{Green} J.~Greensite, ``An introduction to the confinement problem'',\\ Lect. Notes Phys. {\bf 972} (2020) 1.

\end{thebibliography}
\end{document}